\documentclass[conference]{IEEEtran}
\IEEEoverridecommandlockouts

\usepackage{cite}
\usepackage{amsmath,amssymb,amsfonts}
\usepackage{algorithmic}
\usepackage{graphicx}
\usepackage{textcomp}
\usepackage{xcolor}
\usepackage{booktabs}
\usepackage{multicol}
\usepackage{multirow}
\def\BibTeX{{\rm B\kern-.05em{\sc i\kern-.025em b}\kern-.08em
    T\kern-.1667em\lower.7ex\hbox{E}\kern-.125emX}}
\begin{document}

\title{Spectral-Temporal Fusion Representation for Person-in-Bed Detection}
\author{
    \IEEEauthorblockN{
        Xuefeng Yang$^{1,\dagger}$,
        Shiheng Zhang$^{1,\dagger}$\thanks{$\dagger$These authors contributed equally to this work.},
        Jian Guan$^{1,*}$\thanks{*Corresponding author.},
        Feiyang Xiao$^1$,
        Wei Lu$^1$, 
        Qiaoxi Zhu$^2$
        }
    \IEEEauthorblockA{
        $^1$Group of Intelligent Signal Processing (GISP), College of Computer Science and Technology, \\ Harbin Engineering University, Harbin, China\\
        $^2$Acoustics Lab, University of Technology Sydney, Ultimo, Australia\\
        }
}
\maketitle
\begin{abstract}
This study is based on the ICASSP 2025 Signal Processing Grand Challenge's Accelerometer-Based Person-in-Bed Detection Challenge, which aims to determine bed occupancy using accelerometer signals. The task is divided into two tracks: ``in bed" and ``not in bed" segmented detection, and streaming detection, facing challenges such as individual differences, posture variations, and external disturbances. We propose a spectral-temporal fusion-based feature representation method with mixup data augmentation, and adopt Intersection over Union (IoU) loss to optimize detection accuracy. In the two tracks, our method achieved outstanding results of 100.00\% and 95.55\% in detection scores, securing first place and third place, respectively.
\end{abstract}
\begin{IEEEkeywords}
Person-in-bed detection, spectral-temporal fusion, accelerometer signal processing
\end{IEEEkeywords}
\section{Introduction}
\label{sec:1}
Accelerometer-Based Person-in-Bed Detection Challenge\cite{PersonInBedDetectionChallenge}, part of the Signal Processing Grand Challenge at ICASSP 2025, aims to detect bed occupancy using the accelerometer signal embedded within a mattress. This challenge consists of two tracks: Track 1 focuses on classifying pre-segmented signals as ``in bed" or ``not in bed", while Track 2 involves streaming detection, where the goal is to predict bed occupancy at each time frame from continuous accelerometer signals.

The person-in-bed detection task assumes significant differences between the states of ``in bed" and ``not in bed", primarily due to signals related to respiration and heartbeats~\cite{PersonInBedDetectionChallenge}. However, the process is complicated by several factors, such as variations between individuals, differences in positions within ``in bed" signals, external disturbances (e.g., vibrations caused by nearby foot traffic), and motion artifacts from people adjusting their position in bed.

\begin{figure*}
    \centering    \includegraphics[width=.88\textwidth]{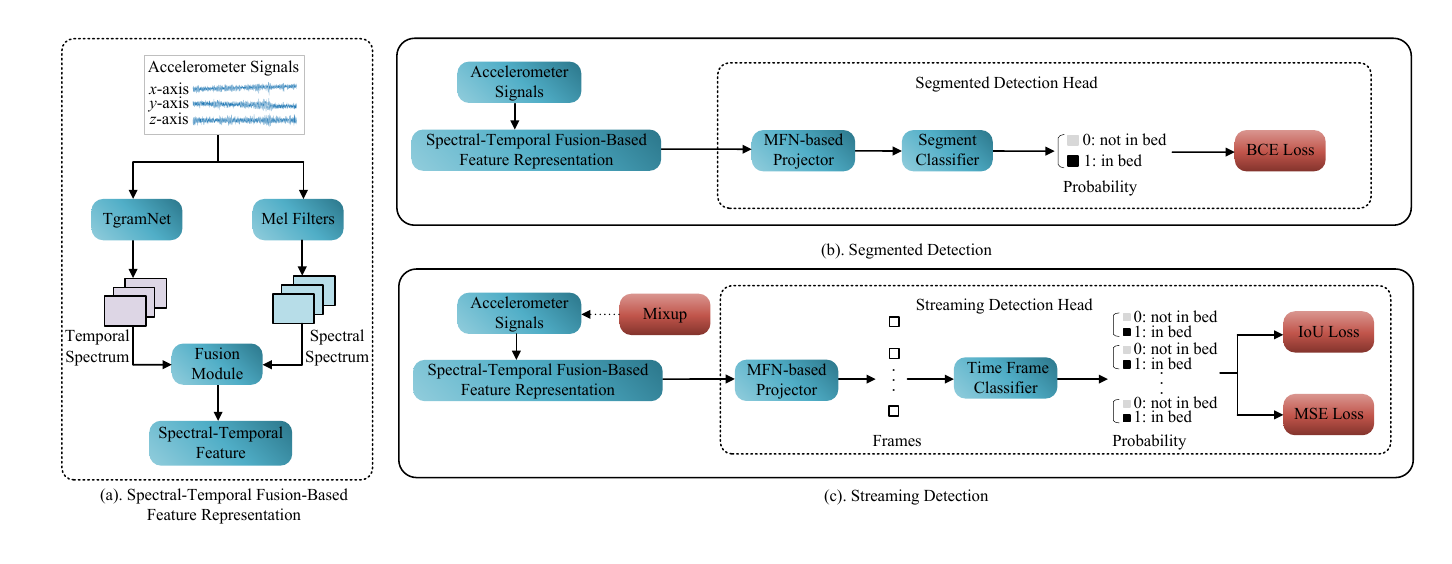}
    \vspace{-8mm}
    \caption{Structure of our models for person-in-bed detection. (a) Spectral-temporal fusion-based feature representation. (b) Segmented detection model and (c) streaming detection model, both based on (a) but operating on different tracks.}
    \label{fig:model}
    \vspace{-5mm}
\end{figure*}

To address these challenges, we adapt the feature representation based on spectral-temporal fusion from our previous work~\cite{LiuSTgram} for Track 1, allowing robust binary classification. Although this strategy was originally used for anomalous sound detection, we modified it for 3-axis accelerometer signals to detect bed occupancy. For Track 2, we extend the detection operation by introducing frame-wise classification for streaming detection, applying mixup data augmentation~\cite{zhang2017mixup} to enhance generalization, and incorporating Intersection over Union (IoU) loss~\cite{Iou} to better align predicted and ground truth event boundaries, thereby improving streaming detection performance. The structure of our proposed feature representation method based on spectral-temporal fusion is given in Fig.~\ref{fig:model}. Our approach achieves a perfect score of 100.00\%, securing first place in Track 1, and a score of 95.55\%, placing third in Track 2.
\section{Method}
\subsection{Spectral-Temporal Fusion-Based Feature Representation}
To differentiate between ``in bed" and ``not in bed" states using accelerometer signals, we employ a spectral-temporal fusion-based feature representation that models person's activity events in both the frequency and time domains. Specifically, the representation is given by:
\begin{equation}
\label{eq:1}
    \mathbf{H} = \mathcal{F}(\mathbf{H}_s, \mathbf{H}_t),
\end{equation}
where $\mathbf{H}_s$ represents the spectral spectrum derived from Mel filters, and $\mathbf{H}_t$ denotes the temporal spectrum obtained from TgramNet~\cite{LiuSTgram}. $\mathcal{F}(\cdot, \cdot)$ is a CNN-based fusion module that combines spectral and temporal information into a single-channel latent feature \(\mathbf{H}\). This fused feature highlights the distinctions between ``in bed" and ``not in bed" states.
\subsection{Segmented Detection Head for Track 1}
For Track 1, which focuses on classifying pre-segmented accelerometer signals, we design a segmented detection head. This module comprises a MobileFaceNet (MFN)-based projector \cite{mfn} and a linear segment classifier. The MFN projector refines the latent spectral-temporal feature $\mathbf{H}$, emphasizing the differences between ``in bed" and ``not in bed" events. The linear segment classifier then predicts the probability of the input segment belonging to each state. The model is optimized using the binary cross-entropy (BCE) loss:
\begin{equation}
    \mathcal{L}_{\text{BCE}} = - \left[ y \log(\mathcal{G}(\hat{\mathbf{H}})) + (1 - y) \log(1 - \mathcal{G}(\hat{\mathbf{H}}))) \right],
\end{equation}
where $y$ denotes the label of the input segment, $y=0$ represents ``not in bed" status, and $y=1$ is ``in bed" status. $\hat{\mathbf{H}}$ denotes the output feature of the MFN projector, and $\mathcal{G}(\hat{\mathbf{H}})$ expresses the predicted possibility for ``in bed" status. This ensures reliable detection of in-bed activities within segmented accelerometer data.
\subsection{Streaming Detection Head for Track 2}
For Track 2, which requires real-time detection, we implement a streaming detection head. Similar to Track 1, this head includes an MFN-based projector and a classifier. However, the classifier operates frame-wise, predicting the probability of ``in bed" for each time frame in the continuous signal. Here, we replace the BCE loss with the mean squared error (MSE) loss to better constrain the prediction for each time frame.

Moreover, to mitigate the influence of the limited training data, we apply the mixup data augmentation technique~\cite{zhang2017mixup} to increase data diversity and improve the generalization of the feature representation. This enhances the model's performance in streaming detection.

In addition, to improve alignment between predicted and ground-truth event boundaries, we incorporate the Intersection over Union (IoU) loss \cite{Iou}. The IoU loss measures the overlap between predicted and ground truth event regions, promoting more accurate localization of ``in bed" event boundaries in the streaming signal.
\section{Results}
\subsection{Dataset and Metric}
Following the Challenge, we conducted experiments using acceleration signals from the ADXL355 accelerometer, collected along the \textit{x}, \textit{y}, and \textit{z} axes at a sampling rate of 250 Hz~\cite{PersonInBedDetectionChallenge}. The evaluation metrics are based on the official scoring criteria set by the competition organizers. For Track 1, the metric measures the accuracy of segment classification. For Track 2, the metric evaluates both the accuracy of event detection and the latency of the predicted event boundaries.
\subsection{Results and Analysis}
Table~\ref{tab:performance} presents a comparison between our models and other state-of-the-art methods for the Accelerometer-Based Person-in-Bed Detection Challenge. Our approach, which utilizes the proposed spectral-temporal fusion-based feature representation, achieved a perfect score of 100.00\% in Track 1, securing the top rank for segmented detection. For Track 2, our model achieves a score of 95.55\%, placing third in streaming detection. These results validate the effectiveness of our approach in handling both segmented and streaming person-in-bed detection tasks.

The strong performance of our models demonstrates the advantages of the spectral-temporal fusion approach, which effectively captures discriminative features for detecting in-bed states. The combination of frame-wise segmentation, data augmentation techniques like mixup, and the integration of Intersection over Union (IoU) loss further enhanced the robustness and accuracy of the streaming detection task.
\begin{table}[t]
    \centering
    \caption{Performance comparison in ICASSP 2025 Accelerometer-Based Person-in-Bed Detection Challenge.}
    \vspace{-2mm}
    \resizebox{\linewidth}{!}{
    \begin{tabular}{ccccc}
        \toprule
        \multirow{2}{*}{Rank} & \multicolumn{2}{c}{Track 1: Segmented Detection} & \multicolumn{2}{c}{Track 2: Streaming Detection} \\
        \cmidrule(lr){2-3} \cmidrule(lr){4-5}
          & Team & Official Score (\%) & Team & Official Score (\%) \\
        \midrule
        1 & \textbf{GISP@HEU} (Ours) & 100.00 & Akhil B & 98.26 \\
        2 & jiayiii gao & 100.00 & SSD-IITM & 98.01 \\
        3 & Akhil B & 99.82 & \textbf{GISP@HEU} (Ours) & 95.55 \\
        4 & SSD-IITM & 99.81 & PES\_DataDelvers & 92.51 \\
        5 & dalest & 99.53 & AdityaRanjanPadhi1 & 91.96 \\
        \bottomrule
    \end{tabular}
    }
    \label{tab:performance}
   \vspace{-4mm}
\end{table}
\section{Conclusion}
In this paper, our proposed model combines spectral and temporal feature representations to tackle challenges in in-bed detection using accelerometer signals. By integrating complementary spectral and temporal features, the model achieves improved detection accuracy. Experimental results show the effectiveness of our approach, achieving strong performance across segmented and streaming detection tracks, showcasing its potential and smart home applications.

\bibliographystyle{IEEEtran}
\bibliography{refs}

\end{document}